\pdfoutput=1

\documentclass[%
 reprint,
 floatfix,
 amsmath,amssymb,
 aps,
]{revtex4-1}

\usepackage{graphicx}
\usepackage{dcolumn}
\usepackage{bm}

\usepackage[utf8]{inputenc}
\usepackage{physics}
\usepackage{amsfonts}
\usepackage{amsmath}

\begin{document}

\preprint{APS/123-QED}

\title{Semi-empirical many-body formalism of optical absorption in nanosystems and molecules.}

\author{Antoine Honet}
\affiliation{%
Department of Physics and Namur Institute of Structured Materials, University of Namur, Rue de Bruxelles 51, 5000 Namur, Belgium
}%

\author{Luc Henrard}
\affiliation{%
Department of Physics and Namur Institute of Structured Materials, University of Namur, Rue de Bruxelles 51, 5000 Namur, Belgium
}%

\author{Vincent Meunier}%
\affiliation{%
 Department of Physics, Applied Physics, and
Astronomy, Rensselaer Polytechnic Institute, Troy, New York 12180, USA,
}%

\date{\today}

\begin{abstract}
A computationally efficient Green's function approach is developed to evaluate the optical properties of nanostructures using a GW formalism applied on top of a tight-binding and mean-field Hubbard model. The use of the GW approximation includes key parts of the many-body physics that govern the optical response of nanostructures and molecules subjected to an external electromagnetic field. Such description of the electron-electron correlation yields data that are in significantly improved agreement with experiments performed on a subset of polycyclic aromatic hydrocarbons (PAHs) considered for illustrative purpose. More generally, the method is applicable to any structure whose electronic properties can be described in first approximation within a mean-field approach and is amenable for high-throughput studies aimed at screening materials with desired optical properties. 
\begin{description}
\item[Keywords]
Tight-binding, Hubbard model, GW approximation, RPA, optical absorption, plasmons, PAH, nano-graphene, quantum plasmonics
\end{description}
\end{abstract}

\maketitle


\author{Antoine Honet, Luc Henrard and Vincent Meunier}

\section{Introduction}
The optical properties of molecules, nanoparticles, and solids are intimately governed by many-body effects, such as electron-electron correlation. Correlation remains, however, difficult to accurately describe in realistic systems, in spite of significant progress in the development of numerical methods and the growing availability of computational resources. The majority of the current theoretical descriptions of optical response of large systems are based on single-electron states, as the treatment of electronic processes are limited to a mean-field approximation due to computational cost. These mean-field approaches include density functional theory (DFT) and the tight-binding (TB) approach. Further, collective phenomena such as plasmons are commonly treated as properties of continuous solid described by a macroscopic dielectric function. They have also been modelled at the quantum level using a number of approaches such as time-dependent density functional theory (TDDFT)~\cite{Runge:1984, Casida:2012} and within the random phase approximation (RPA). These formalisms contribute to a better description of the Coulomb interaction~\cite{Delerue:Nanostructures, Delerue:2017, deAbajo:2013, deAbajo:2015}.  However, methods such as TDDFT are computationally demanding and usually intractable for most realistic system sizes. 

Here, we introduce a new semi-empirical many-body approach that allows for considerably reducing the computational cost of RPA. The method is based on the GW approximation as developed recently in Ref.~\onlinecite{Joost:2019} in the context of the study of magnetic properties of graphene nanoribbons. The method does not require the explicit calculation of the sum over states as formulated in the Lindhard equation since all summations are performed in the frequency domain, thus making it possible to significantly reduce the numerical cost. Moreover, the method presented here focuses on spin correlation that is neglected in RPA. Spin correlation is essential in the description of the electromagnetic response of molecules and solids, especially for open-shell systems that feature unpaired electrons. The Hubbard model has been developed to include such effects and, in its original formulation, takes many-body interactions into account~\cite{Hubbard:1963}. However, it is usually implemented within a mean-field approximation. Here we move towards a many-body approach with the development of a GW correction to this mean-field approximation, using a TB Hamiltonian as a starting point. 

We illustrate the accuracy of the new approach for the evaluation of the optical response of polycyclic aromatic hydrocarbons (PAHs). In a way, these molecules can be seen as the smallest nano-graphene systems. Interestingly, they have been shown to host collective excitations (\textit{e.g.}, molecular plasmons) in the near-infrared and visible range~\cite{deAbajo:2013}. They have been recently investigated both theoretically and experimentally, leading to a proof-of-concept of an electrochromic device~\cite{deAbajo:2015}.  The results presented here for PAHs highlight the importance of many-body effects, especially in the case of open-shell systems such as PAH anions.

However, the proposed formalism and its numerical implementation are not restricted to a particular type of molecules or nanosystems, so long as the validity of a TB description as a starting point holds. For example, quantum dots of semiconductors ~\cite{Delerue:Nanostructures}, nanoparticles of metal oxide such as ZnO ~\cite{Delerue:2017}, or pristine or defective 2D materials~\cite{CastroNeto:2009} are also amenable to the theoretical analysis performed here.  

The rest of the paper is organized as follows: We first present the model for the optical absorption cross-section both using Lindhard formula and the RPA approach (Section II.A) and in the Green's function formalism (Section II.B). We then introduce the Hubbard model and its diagonalization in the mean-field approximation as well as in the GW approximation (Section II.C). In section III, we numerically investigate the optical properties of several PAH molecules. We compare the calculated results with experimental data from Ref.~\onlinecite{deAbajo:2015} and discuss the merit of different levels of approximation: TB, mean-field approximation of the Hubbard model (MF-H), and GW correction to include many-body effects on top of MF-H model (GW-H). The examples illustrate the success of the two-parameter approach developed here by showing how it reproduces salient experimental features of the optical response of the molecules. More importantly, we demonstrate that the explicit inclusion of correlation, beyond the mean-field approach, can be performed at fairly low computational cost and is crucial for the computational design of optically active material systems. 

\section{Model and methods}
\subsection{\label{sec:level1} Tight-Binding approximation}
\subsubsection{Tight-binding Hamiltonian}
The tight-binding approximation is a ubiquitous mean-field approach used for the description of the electronic properties of molecules and solids~\cite{Delerue:Nanostructures}. In the TB framework, the Hamiltonian reads
\begin{equation}
    \hat{H}_{TB} = \sum_{i,l,\sigma} \epsilon_{i,l,\sigma} \hat{c}^\dagger_{i,l\sigma} \hat{c}_{i,l\sigma} - \sum_{<ij>,ll' \sigma} t_{ij,ll'}^\sigma \hat{c}^\dagger_{i,l\sigma} \hat{c}_{j,l'\sigma} + cc.
\label{TB_Hamiltonian}
\end{equation}
where indices $i$ and $j$ refer to the atomic sites, $\sigma$ is the spin of the electron, $l$ and $l'$ are the indices of the orbitals, $t_{ij,ll'}^\sigma$ are the hopping parameters, and $\epsilon_{i,l,\sigma}$ are the on-site energy parameters. $\hat{c}^\dagger_{i,l\sigma}$ and $\hat{c}_{i,l\sigma}$ are the creation and annihilation operators of an electron in a state $\ket{il\sigma}$. The second sum marked by "$<\ldots>$" in Eq.~\ref{TB_Hamiltonian} is restricted to shells of nearest-neighbors.

Diagonalization of the Hamiltonian yields the eigenenergies $E_k$ and the coefficients $a^k_{il\sigma}$ of the eigenstates $ \ket{\Psi_k}$: 
\begin{equation}
    \ket{\Psi_k} = \sum_{i, l ,\sigma} a^k_{il\sigma} \hat{c}^\dagger_{il\sigma} \ket{0}
    \label{TB_states}
\end{equation}
where $\ket{0}$ is the state with no electron.

\subsubsection{Lindhard formula and random phase approximation}
From the eigen-states of the Hamiltonian (Eq.~\ref{TB_states}), the non-interacting atomic susceptibility can be computed from Lindhard formula~\cite{Delerue:Nanostructures, Delerue:2017, Hedin:1969}:
\begin{equation}
    \chi^0_{i\sigma, j\sigma '} (\omega) = \sum_{m, m', l, l'} \frac{(f_m-f_{m'}) a_{il\sigma}^m (a_{il\sigma}^{m'})^* a_{jl'\sigma'}^m (a_{jl'\sigma'}^m{'})^*}{\hbar \omega - (E_{m'}-E_m)+i\zeta }
\label{Lind_form}
\end{equation}
where $f_m = f_{FD}(E_m)$ is the Fermi-Dirac distribution and $\zeta$ is a small positive real number. 

The interacting susceptibility in the random phase approximation includes the Coulomb interaction between electronic transitions and is given by~\cite{Delerue:Nanostructures, Delerue:2017, Hedin:1969}:
\begin{equation}
    \chi^{RPA} = \chi^0 \epsilon^{-1}
\label{chi_RPA}
\end{equation}
where $\epsilon = I -V\chi^0$ is the dielectric matrix and $V$ is the Coulomb matrix ($e^2/R$ except for the diagonal elements that are computed from the electron density probability~\cite{Thongrattanasiri:2012}). Eq.~\ref{chi_RPA} must be read as a matrix equation, expressed in the tight-binding basis. 

The absorption cross-section ($\sigma_{abs} (\omega)$) of a system subjected to a uniform electric field $\Vec{E}_{ext}(\omega)$ can be expressed as~\cite{Delerue:2017, Hedin:1969}:
\begin{equation}
    \sigma_{abs} (\omega) = 4\pi \frac{\omega}{c} Im(\alpha(\omega))
\label{sigma_abs}
\end{equation}
where $c$ is the speed of light and $\alpha(\omega)$ is the polarizability given by~\cite{Delerue:2017, Hedin:1969}:
\begin{equation}
    \alpha(\omega) = -e^2\sum_{i\sigma,j\sigma'} \chi^{RPA}_{i\sigma, j\sigma'} (\Vec{R}_i \cdot \Vec{u}) (\Vec{R}_j \cdot \Vec{u})
\label{alpha}
\end{equation}
in terms of $\Vec{R}_i$  and  $\Vec{u}$, which are the position of atom $i$ and the unitary vector along the direction of the external electric field, respectively.

\subsection{Green's function formalism}
 The non-interacting susceptibility $\chi_0$ (Eq.~\ref{Lind_form}) can be evaluated in the Fourier space within the Green function formalism~\cite{Joost:2019} at a much reduced computational cost. We note that a similar method was used in Ref.~\onlinecite{Thongrattanasiri:2012}, although these authors did not explicitly refer to their approach as a Green's function formalism. 
\begin{figure*}[ht]
\centering
    \includegraphics[width=\textwidth]{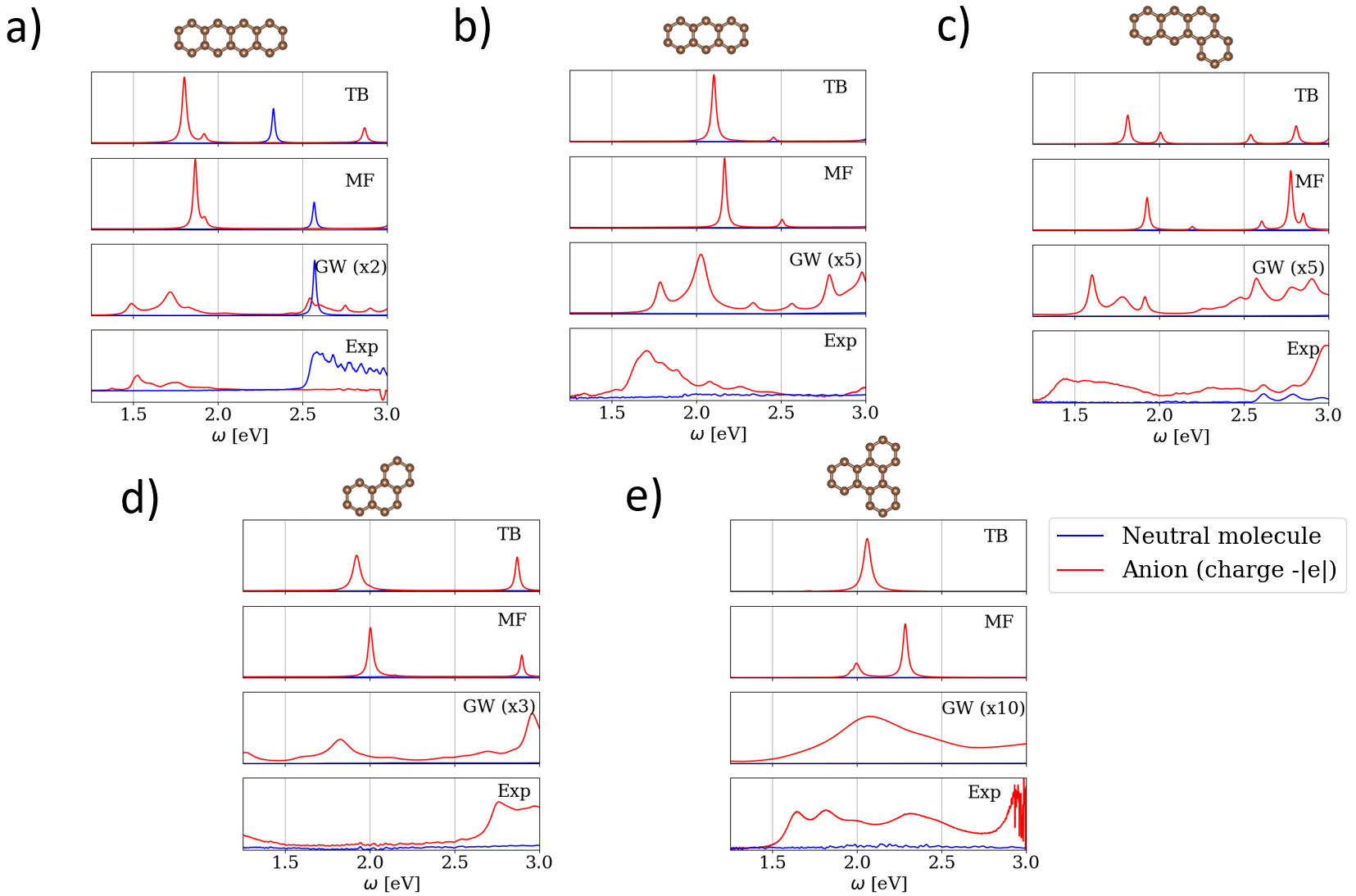}
\caption{Optical absorption cross-section computed within TB (upper panels), MF-H (second panels from the top) and GW-H (third panels from the top) approaches for tetracene (a), anthracene (b), tetraphene (c), phenanthrene, (d) and triphenylene (e). Experimental results from Ref.~\onlinecite{deAbajo:2015} (lower panel) are displayed for comparison. For the simulations, the average is taken on the orientations of the electric field in the plane of the molecule. The cross-section are given in arbitrary units (the same for all the theoretical curves except the GW-H curves that have been rescaled (the scaling factors are shown on the plots). The parameters used to compute the results are $U=2t$, $t=2.6$~eV, and $\eta = 2.10^{-4}\frac{E_H}{t}$ (with $E_H=27.21$~eV, the Hartree energy).}
\label{fig:All_compar}
\end{figure*}

\subsubsection{Green's function and non-interacting susceptibility}
We will adopt the usual representation of Green's functions as matrices expressed in the TB orbital basis. We first define the retarded (R), advanced (A) Green's functions as:
\begin{equation}
    G^{R/A}_{il\sigma, jl'\sigma'} (\omega) = \bra{il\sigma} (\hbar\omega - \hat{H} \pm i\eta)^{-1} \ket{jl'\sigma'} 
\end{equation}
where $\hat{H}$ is the Hamiltonian and $\eta$ is a small positive real number. Connection between $\eta$ and $\zeta$ (Eq.~\ref{Lind_form}) is explored in the Supplementary Information.

The lesser ($<$) and greater ($>$) Green's functions can be expressed in terms of the $R/A$ Green's functions by the following relationship:
\begin{equation}
\begin{split}
    &G^<(\omega) = -f_{FD}(\omega-\mu) [G^R(\omega)-G^A(\omega)]\\
    &G^>(\omega) = \bar{f}_{FD}(\omega-\mu) [G^R(\omega)-G^A(\omega)]
\end{split}
\end{equation}
with $\bar{f}_{FD}(\omega) = 1-f_{FD}(\omega)$.

The non-interacting susceptibility expressed in time domain is calculated using the following formulate~\cite{Joost:2019}:
\begin{equation}
    \chi^0(t) = -i\hbar \Theta(t) [G^>(t)\circ (G^<(t))^* - G^<(t)\circ (G^>(t))^*]
\label{chi0_time}
\end{equation}
where $\Theta(t)$ is the Heaviside step-function. In this expression, the Green's functions have been Fourier transformed and $\circ$ denotes the Hadamard product (\textit{i.e.}, element-wise product) between matrices. Finally, the non-interacting susceptibility in frequency space is found by Fourier transforming Eq.~\ref{chi0_time}.

\subsection{Hubbard model}
Clearly, only one-body operators (hopping terms) are considered in the tight-binding Hamiltonian of Eq.~\ref{TB_Hamiltonian}. This corresponds to a mean-field approach where electrons are treated as independent particles evolving in the mean-field potential due to all the other particles. Moving beyond the mean-field approach, the interactions between electrons of opposite spin on the same site can be turned on using the Hubbard model~\cite{Hubbard:1963}:
\begin{equation}
\begin{split}
    \hat{H}_{Hubbard} &= \hat{H}_{TB} +  \sum_{i, l} U_{il} \hat{n}_{il\sigma} \hat{n} _{il\sigma'}
\end{split}
\label{Hubbard_Hamiltonian}
\end{equation}
where $\hat{n}_{il\sigma} = \hat{c}^\dagger_{il\sigma} \hat{c}_{il\sigma}$ is the density operator for particles with spin $\sigma$, in the orbital $l$, and located on site $i$. $U_{il}$ are interaction parameters. The interaction term (the second term in Eq.~\ref{Hubbard_Hamiltonian}) accounts for on-site correlations. It contains a 2-body operator that makes the problem very difficult to solve exactly because the Hilbert space grows exponentially with the number of electrons~\cite{Sharma:2015}.

\subsubsection{Mean-field and GW approximation of the Hubbard model}
The most common approximation made to obtain tractable solutions to the Hubbard model is to treat the 2-body operators in a mean-field approximation
~\cite{Girao:2015,Yazyev:2010}:
\begin{equation}
    H_{Hub, MF} = - \sum_{ij, \sigma} t_{ij} c^\dagger_{i\sigma} c_{j\sigma} +  \sum_{il} U_{il} (\hat{n}_{il\sigma} \langle \hat{n}_{il\sigma'} \rangle + \langle \hat{n}_{il\sigma} \rangle \hat{n}_{il\sigma'} ).
\label{Hubbard_MF}
\end{equation}
In this equation, $\langle . \rangle$ represents the mean value of the density operator.

This Hamiltonian has to be solved self-consistently since the average occupations can only be known after the eigenstates have been determined. This means that the Hamiltonian is diagonalized at step $k+1$ based on the mean occupations at step $k$ ($\langle \hat{n}^{k}_{il\sigma} \rangle$) until convergence is reached. This MF-H model has been used extensively in the context of graphene physics and has yielded results in very good agreement with experiment~\cite{Girao:2015, CastroNeto:2009, Yazyev:2010}.

An important improvement, beyond the mean-field approach, can be obtained by considering GW corrections to the mean-field solutions (GW-H model) ~\cite{Hedin:1969, Joost:2019, Onida:2002, Aryasetiawan:1998} and by using the Green's function in a self-consistent procedure as proposed in Ref.~\onlinecite{Joost:2019}. The algorithm is based on Hedin's equations~\cite{Hedin:1969}. The main idea is to first solve the MF-H Hamiltonian (Eq.~\ref{Hubbard_MF}) and to set the initial, non-interacting (subscript $0$) Green's functions as:
\begin{equation}
    G^{R/A}_{0, il\sigma, jl'\sigma'} (\omega) = \bra{il\sigma} (\hbar\omega - \hat{H}_{Hub,MF} \pm i\eta)^{-1} \ket{jl'\sigma'}.
\end{equation}
Dyson equation is used to calculate the corrected Green's functions:
\begin{equation}
    G^{R/A} (\omega) = G^{R/A}_0 (\omega) + G^{R/A}_0 (\omega) \Sigma^{R/A} (\omega) G^{R/A} (\omega).
\label{Dyson_equ}
\end{equation}
Equivalently, Dyson equation can also be written as:
\begin{equation}
    [G^{R/A} (\omega)]^{-1} = [G^{R/A}_0 (\omega)]^{-1} - \Sigma^{R/A} (\omega).
\label{Dyson_equ_inv}
\end{equation}
If the self-energies $\Sigma^{R/A} (\omega)$ were exactly known, we could find the exact solution for $G^{R/A} (\omega)$ from $G^{R/A}_0 (\omega)$. In practice, the self-energies are approximated and in the GW approximation, their expressions are given by:
\begin{equation}
    \Sigma^{\lessgtr} (t) = i\hbar W^{\lessgtr} (t)~\circ G^{\lessgtr} (t)
\end{equation}
where $W(\omega)$ is the dynamically-screened potential derived from $\chi^0(\omega)$ and a Dyson-like equation~\cite{Stefanucci:Book}:
\begin{equation}
    W^R(\omega) = V + V \chi^0(\omega) W^R(\omega),
\label{screened_W}
\end{equation}
where $V$ is now the potential matrix, containing the $U$-term of the Hubbard Hamiltonian (Eq.~\ref{Hubbard_Hamiltonian}) as diagonal terms in space and off-diagonal terms in spin. 

The relations between the $R/A$ and $\lessgtr$ Green's functions for the screened potential and the self-energy are given by:
\begin{equation}
    \begin{split}
        &W^>(\omega) = 2 f_{BE} (\omega - \mu) Im(W^R(\omega))\\
        &W^<(\omega) = 2 (f_{BE}(\omega - \mu)+1)  Im(W^R(\omega))\\
        &\Sigma^R(t) = \Theta(t) [\Sigma^>(t) - \Sigma^<(t)]
    \end{split}
\end{equation}
where $f_{BE} (\omega - \mu)$ is the Bose-Einstein distribution.

Using the description of TB, MF-H and GW-H in the Green's function formalism, we can now compute the susceptibility for each model using Eq.~\ref{chi0_time} and then apply the RPA (Eq.~\ref{chi_RPA}). It is then straightforward to compute the polarizability and the absorption cross-section of a uniform electric field using Eqs.~\ref{sigma_abs} and \ref{alpha}.

We note that the originality of the present approach is that GW is here applied on top of the MF-H model rather than on top of DFT, as it is usually done. In addition, as an extension of the work of Ref.~\onlinecite{Joost:2019}, which introduced this approach to calculate STM images of graphitic nanoribbons, we employ this method to evaluate optical properties. We will illustrate this approach in the next section for the case of selected PAHs.

\section{Application to polycyclic aromatic hydrocarbons}
\subsection{Single-band model Hamiltonians}
Polycyclic aromatic hydrocarbons (PAHs) are small molecules made of carbon and hydrogen atoms organized in aromatic cycles. They can be considered as graphene nano-dots. Nanostructured graphene as well as PAHs are $\pi$-conjugated materials that are well described by a single-band model (the $p_z$ orbitals of carbon) in the low-energy regime~\cite{CastroNeto:2009}. Similar to most previous studies devoted to PAHs or graphene~\cite{deAbajo:2013, Yazyev:2010, CastroNeto:2009}, we will only consider first nearest-neighbour hopping terms and we will posit that the parameters $t_{ij}$ and $U_{i}$ are the same for all atomic sites: $t_{ij}=t$, $U_{i}=U$. We also assume that all carbon atoms have the same on-site energies and choose it equal to zero: $\epsilon_{i,l,\sigma}=0$, thus fixing the Fermi energy of the TB spectrum at $E_F=0$. In all PAHs, edge carbons are considered as passivated by hydrogen atoms and all the carbon atoms are therefore sp$^2$-bounded with no dangling bonds.  

A hopping parameter of $t=2.6$~eV  is a typical value adopted to describe graphene-like structures~\cite{Yazyev:2010} and has been used both in TB and Hubbard models. In the Hubbard model, we have investigated a large range of $U$ values and found that $U=2t$ reproduces better the experimental data, as we will show below. We will use this value unless otherwise stated. We note that the actual value of $U$ is still debated in the literature as different authors have employed parameters ranging from less than $t$~\cite{Girao:2015} to as much as $4t$~\cite{CastroNeto:2009}, depending on the details of the structures considered.

 Experiments indicate that molecular plasmons in PAHs can be tuned by charging the molecule with electrons~\cite{deAbajo:2015}. For this reason, we have computed the absorption cross-section for neutral and single-electron charged molecules in the three models considered in this work. We have considered the five PAH molecules for which experimental data are available (Fig.~\ref{fig:All_compar})~\cite{deAbajo:2015}. 

\subsection{Numerical results}
\subsubsection{Tetracene}
This section presents a detailed analysis of our computational approach applied to tetracene. Fig.~\ref{fig:All_compar} (a) compares the optical absorption cross-section computed from Eq.~\ref{sigma_abs} using TB, MF-H, and GW-H approaches with the experimental data~\cite{deAbajo:2015}. Inspection of the plots indicate that the GW-H model reproduces better the main features of the experimental data. This can be seen by examining how well the experimental positions and the shape of the peaks between $1.5$~eV and $2$~eV (in the one electron charged case) and the resonance (in the case of the neutral system) are reproduced.

\begin{figure}
\centering
    \includegraphics[width=8cm]{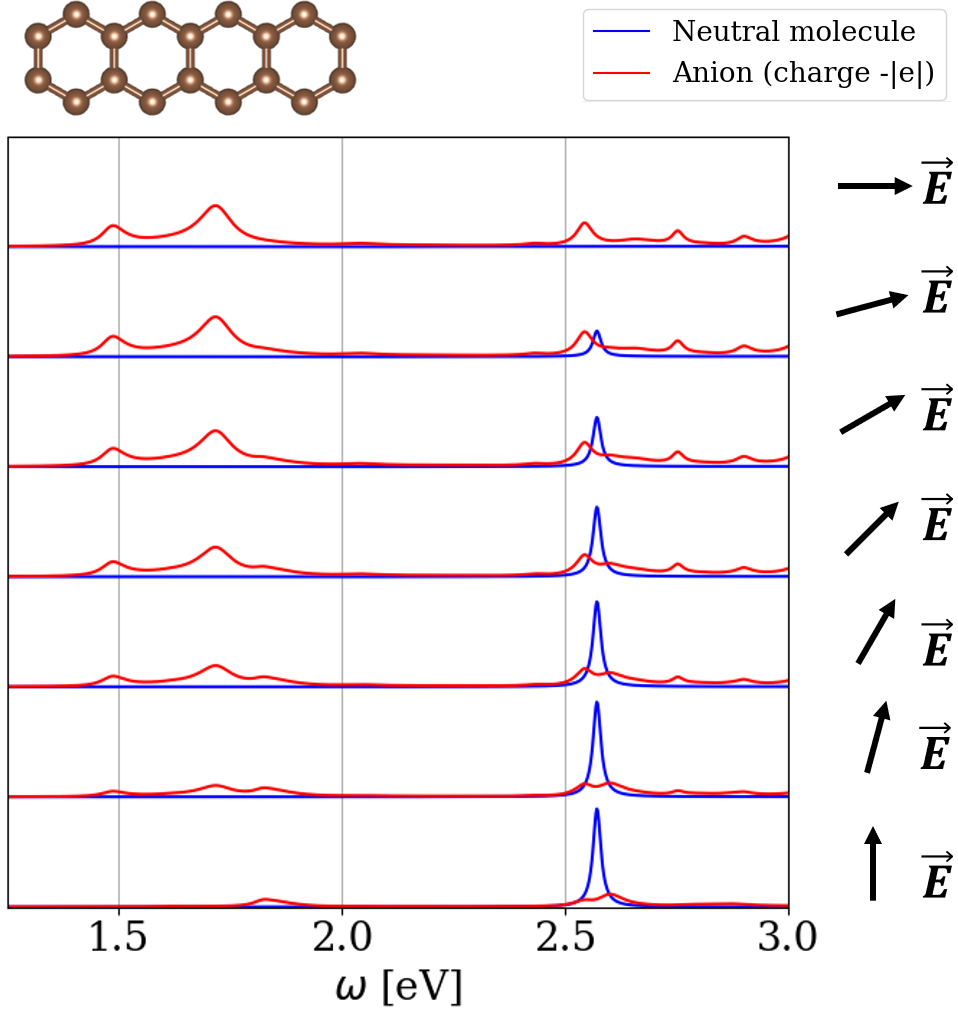}
\caption{Optical absorption cross-section of tetracene in the neutral state (blue curves) and in the one-electron charged state (red curves) in the GW-H model for different polarizations of the incident electric field (see arrows, angles are spaced by $\frac{\pi}{12}$). The parameters used to compute the data are $U=2t$, $t=2.6$~eV, and $\eta = 2.10^{-4}\frac{E_H}{t}$.}
\label{fig:tetracene_angles_GW_RPA}
\end{figure}

The computed spectra shown in Fig.~\ref{fig:All_compar} are taken as an average over the polarization of the incident electric field in the plane of the molecule. Fig.~\ref{fig:tetracene_angles_GW_RPA} illustrates the absorption cross-section computed in the GW-H approximation for different orientations of the incident electric field. As expected, we observe that the relative intensities of the peaks change with the polarization but not their energies. The low-energies features for the charged molecule ($1.5$~eV to $2$~eV) reach a maximum for a polarization parallel to the main axis of the molecule where the absorption above $2.5$~eV dominates for the perpendicular polarization for neutral systems. The absorption cross-section for the TB and MF-H models also strongly depends on the incident electric field polarization (See Supplementary Information).

\begin{figure}[h!]
\centering
    \includegraphics[width=0.48\textwidth]{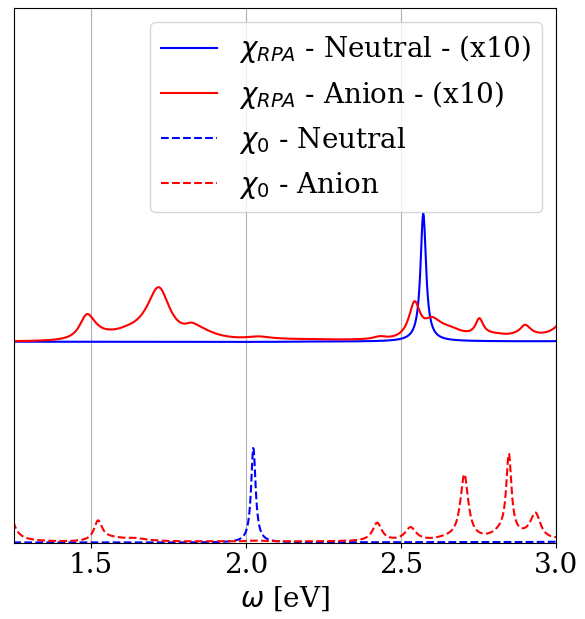}
    
\caption{Optical absorption cross-section of the tetracene molecule in the neutral state (blue curves) and in the one-electron charged state (red curves) within the GW-H approximation without RPA screening (computed with $\chi_0$) (dashed lines) and with RPA screening (solid lines). The units are arbitrary and the RPA curves are multiplied by a factor $10$. The RPA and non-interacting curves have been shifted for better visualization. The values of the parameters are $U=2t$, $t=2.6 eV$ and $\eta = 2.10^{-4}\frac{E_H}{t}$.}
\label{fig:tetracene_MF_RPA_vs_chi0}
\end{figure}

We now present a more detailed analysis of the importance of the RPA screening and of the strength of the on-site electron correlation (parameter $U$) on the simulated absorption cross-section. Fig.~\ref{fig:tetracene_MF_RPA_vs_chi0} shows the absorption cross-section computed with and without the RPA correction in the GW-H model. The importance of the screening due to the Coulomb potential in the RPA is demonstrated by the significant differences between absorption calculated with $\chi_0$ (no screening) and with $\chi_{RPA}$ (screening included). For the neutral tetracene molecule, the main transition in the considered energy range is blueshifted by more than 0.5 eV when the Coulomb screening is included. For the tetracene anion, the susceptibility is totally modified with RPA corrections. The effect of RPA screening in the MF-H approximation is presented in the Supplementary Information. It has also been extensively studied in Ref.~\onlinecite{deAbajo:2013} within TB. The absorption cross-section computed with the non-interacting $\chi_0$ is the signature of the single electron-hole transitions whereas RPA includes the Coulomb interaction between these transitions. RPA thus describes more accurately plasmonic (collective) phenomena (see, \textit{e.g.}, Ref.~\onlinecite{deAbajo:2013}). 

We emphasize here that RPA has to be considered in order to include  screening \textit{via} the full Coulomb potential. The GW-H calculations include interaction between electrons of opposite spin on the same site and the RPA includes the inter-site Coulomb effects on the top of the spin interactions.

\begin{figure}
\centering
    \includegraphics[width=0.5\textwidth]{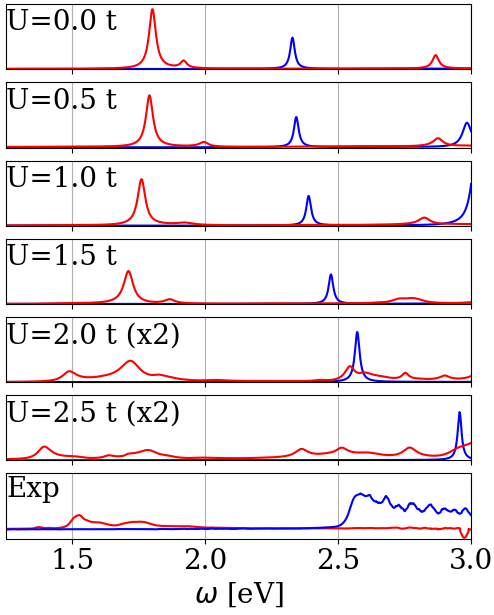}
\caption{Optical absorption cross-section of tetracene in the neutral state (blue curves) and in the one-electron charged state (red curves) within the GW-H model for different values of $U$ ranging from $U=0$ to $U=2.5 t$ (top pannels). Experimental results from Ref.~\onlinecite{deAbajo:2015} are shown for comparison (bottom panel). For the simulations, an average over incident field orientations is performed. The cross-sections are given in arbitrary units that are the same for the theoretical curves except for $U=2t$ and $U=2.5t$ curves that have been multiplied by a factor of $2$ for better visualization. The values of the other parameters used here are $t=2.6 eV$ and $\eta = 2.10^{-4}\frac{E_H}{t}$.}
\label{fig:tetracene_differents_U}
\end{figure}

Up to now, we have presented results using $U=2t$. In Fig.~\ref{fig:tetracene_differents_U}, we investigate a range of $U$ values to highlight their effects on the absorption cross-section of tetracene in the GW-H model. We also compare the simulated spectra with the experimental data. As $U$ increases, the low-energy peaks of the one-electron charged tetracene molecule are redshifted but also split into many substructures.  At $U=2t$, the positions of the two main low-energy peaks are in good agreement with the experimental observation. When considering a larger value of $U$ such as $U=2.5t$, the lowest energy state redshifts further towards energies smaller than the experimental ones. In contrast, the main peak of the neutral molecule blueshifts when $U$ grows from $U=0$ to $U=2.5t$. At $U=2t$, the energy of this peak matches that of the experiment.

We have performed the same systematic study of the role of the parameter $U$ on the optical absorption cross-section in the MF-H model (See Supporting Information). We found that the mean-field approximation does not reproduce the experimental data in a satisfactory manner for any value of $U$, in clear contrast to the GW approximation.

\subsubsection{Comparison with experiment for other PAHs}
A comparison of the absorption cross-section for four other PAHs that were experimentally studied in Ref.~\onlinecite{deAbajo:2015} is presented in Fig.~\ref{fig:All_compar}. The comparison further demonstrates that the GW-H model is in much better agreement with experiment compared to the mean-field approximations. This emphasizes the important role on-site correlations play in the optical properties of carbon clusters. It also indicates that the optical properties of PAHs are well described by the one-orbital Hubbard model. The same values of the parameters ($t=2.6 eV$ and $U=2t$) have been used for all the molecules, thereby suggesting that these parameters can also be employed to describe larger PAHs or nano-graphene clusters.

For anthracene (fig.~\ref{fig:All_compar} (b)), the low-energy peaks for the charged molecules are redshifted towards the experimental energies when the GW approximation is used. We also note the emergence of satellite peaks in the GW-H spectrum that seem to correspond to experimental features. When considering the charged tetraphene (fig.~\ref{fig:All_compar} (c)), we see the emergence of a group of peaks between $1.5$~eV and $2$~eV, visible as a continuum in the experimental spectrum. In addition, a minimum of absorption is observed both in GW-H and experimentally around $2$~eV and another group of peaks is present between $2.5$~eV and $3$~eV.

For phenanthrene (fig.~\ref{fig:All_compar} (d)), one effect of the GW corrections is to lower the position of the peak that is just below $2$~eV in TB and MF (note that those peaks are not present in the experimental spectrum). Furthermore, there is a new small peak around $2.7$~eV in the GW-H spectrum, corresponding to the one seen in the experiment spectrum. Note, finally, the presence of another peak around $2.9$~eV that is more pronounced in GW-H than in TB and MF-H as well as in experiment.

Triphenylene (fig.~\ref{fig:All_compar} (e)) constitutes a less convincing example of match between theory and experiment. We note, however, that the absorption spectrum is wider in GW-H than in TB or MF-H and we can see some multiple peak features a that are also present in the experimental spectrum. Overall, looking at fig.~\ref{fig:All_compar}, the agreement with experimental data is well improved for all the PAHs considered when applying the GW approximation.

\section{Conclusion}
We introduced a Green's function formalism that includes electron-electron spin correlation based on a GW approximation of the Hubbard model within a TB-based approach. We compute the optical cross-section of PAHs and demonstrate the role of spin correlation by comparing with TB and mean-field methods. The Green's function formalism also reduces the computational cost associated with the evaluation of the non-interacting susceptibility ($\chi_0$) usually given by the Lindhard formula, thus leading to the possibility of investigating larger and more complex systems.

The use of GW and inclusion of correlation on top of a tight-binding approach is slated to play an important role in quantum plasmonics because it deals with collective excitations that are likely to be strongly dependent on correlation. These correlations are expected to play an important role especially in open-shell systems or in topological states of graphene nanoribbons~\cite{Joost:2019}, among other systems. The efficiency of our approach also opens the possibility to investigate more complex systems than can be described by semi-empirical methods such as defective 2D materials and doped semiconductor nanocrystals.

\section*{Acknowledgements}
We thank the authors of Ref.~\onlinecite{deAbajo:2015} for providing us the published experimental data reproduced on the figures of the present publication. V.M. acknowledges the Francqui foundation for the support during his stay at UNamur as International Francqui Professor 2018-2019. A.H. is a Research Fellow of the Fonds de la Recherche Scientifique - FNRS. This research used resources of the "Plateforme Technologique de Calcul Intensif (PTCI)" (http://www.ptci.unamur.be) located at the University of Namur, Belgium, and of the Université catholique de Louvain (CISM/UCL) which are supported by the F.R.S.-FNRS under the convention No. 2.5020.11. The PTCI and CISM are member of the "Consortium des Équipements de Calcul Intensif (CÉCI)" (http://www.ceci-hpc.be).
\bibliographystyle{plain}
\bibliography{references}

\begin{thebibliography}{10}

\bibitem{Amendola:2017}
Vincenzo Amendola, Roberto Pilot, Marco Frasconi, Onofrio~M Marag{\`{o}}, and
  Maria~Antonia Iat{\`{\i}}.
\newblock Surface plasmon resonance in gold nanoparticles: a review.
\newblock {\em Journal of Physics: Condensed Matter}, 29(20):203002, apr 2017.

\bibitem{Aryasetiawan:1998}
F~Aryasetiawan and O~Gunnarsson.
\newblock {TheGWmethod}.
\newblock {\em Reports on Progress in Physics}, 61(3):237--312, mar 1998.

\bibitem{Baburin:2019}
Aleksandr~S. Baburin, Alexander~M. Merzlikin, Alexander~V. Baryshev, Ilya~A.
  Ryzhikov, Yuri~V. Panfilov, and Ilya~A. Rodionov.
\newblock Silver-based plasmonics: golden material platform and application
  challenges [invited].
\newblock {\em Opt. Mater. Express}, 9(2):611--642, Feb 2019.

\bibitem{Girao:2015}
Zachary Bullard, Eduardo~Costa Gir{\~a}o, Jonathan~R. Owens, William~A.
  Shelton, and Vincent Meunier.
\newblock Improved all-carbon spintronic device design.
\newblock {\em Scientific Reports}, 5(1):7634, Jan 2015.

\bibitem{Casida:2012}
M.E. Casida and M.~Huix-Rotllant.
\newblock Progress in time-dependent density-functional theory.
\newblock {\em Annual Review of Physical Chemistry}, 63(1):287--323, 2012.
\newblock PMID: 22242728.

\bibitem{CastroNeto:2009}
A.~H. Castro~Neto, F.~Guinea, N.~M.~R. Peres, K.~S. Novoselov, and A.~K. Geim.
\newblock The electronic properties of graphene.
\newblock {\em Rev. Mod. Phys.}, 81:109--162, Jan 2009.

\bibitem{deAbajoCox:2014}
Joel~D. Cox and F.~Javier Garc{\'i}a~de Abajo.
\newblock Electrically tunable nonlinear plasmonics in graphene nanoislands.
\newblock {\em Nature Communications}, 5(1):5725, Dec 2014.

\bibitem{Delerue:Nanostructures}
C.~Delerue and M.~Lannoo.
\newblock {\em Nanostructures: Theory and Modeling.}
\newblock Springer Berlin Heidelberg., 2004.

\bibitem{Delerue:2017}
Christophe Delerue.
\newblock Minimum line width of surface plasmon resonance in doped zno
  nanocrystals.
\newblock {\em Nano Letters}, 17(12):7599--7605, 2017.
\newblock PMID: 29190107.

\bibitem{Feldner:2010}
H\'el\`ene Feldner, Zi~Yang Meng, Andreas Honecker, Daniel Cabra, Stefan
  Wessel, and Fakher~F. Assaad.
\newblock Magnetism of finite graphene samples: Mean-field theory compared with
  exact diagonalization and quantum monte carlo simulations.
\newblock {\em Phys. Rev. B}, 81:115416, Mar 2010.

\bibitem{deAbajo:2014}
F.~Javier García~de Abajo.
\newblock Graphene plasmonics: Challenges and opportunities.
\newblock {\em ACS Photonics}, 1(3):135--152, 2014.

\bibitem{Hedin:1969}
Lars Hedin and Stig Lundqvist.
\newblock Effects of electron-electron and electron-phonon interactions on the
  one-electron states of solids.
\newblock volume~23 of {\em Solid State Physics}, pages 1 -- 181. Academic
  Press, 1970.

\bibitem{Hubbard:1963}
J.~Hubbard.
\newblock Electron correlations in narrow energy bands.
\newblock {\em Proceedings of the Royal Society of London. Series A,
  Mathematical and Physical Sciences}, 276(1365):238--257, 1963.

\bibitem{Jiang:2018}
Nina Jiang, Xiaolu Zhuo, and Jianfang Wang.
\newblock Active plasmonics: Principles, structures, and applications.
\newblock {\em Chemical Reviews}, 118(6):3054--3099, 2018.
\newblock PMID: 28960067.

\bibitem{Joost:2019}
Jan-Philip Joost, Antti-Pekka Jauho, and Michael Bonitz.
\newblock Correlated topological states in graphene nanoribbon
  heterostructures.
\newblock {\em Nano Letters}, 19(12):9045--9050, 2019.
\newblock PMID: 31735027.

\bibitem{deAbajo:2015}
Adam Lauchner, Andrea~E. Schlather, Alejandro Manjavacas, Yao Cui, Michael~J.
  McClain, Grant~J. Stec, F.~Javier García~de Abajo, Peter Nordlander, and
  Naomi~J. Halas.
\newblock Molecular plasmonics.
\newblock {\em Nano Letters}, 15(9):6208--6214, 2015.
\newblock PMID: 26244925.

\bibitem{Li:2017}
Yu~Li, Ziwei Li, Cheng Chi, Hangyong Shan, Liheng Zheng, and Zheyu Fang.
\newblock Plasmonics of 2d nanomaterials: Properties and applications.
\newblock {\em Advanced Science}, 4(8):1600430, 2017.

\bibitem{Lindquist:2012}
Nathan~C Lindquist, Prashant Nagpal, Kevin~M McPeak, David~J Norris, and
  Sang-Hyun Oh.
\newblock Engineering metallic nanostructures for plasmonics and nanophotonics.
\newblock {\em Reports on Progress in Physics}, 75(3):036501, feb 2012.

\bibitem{deAbajo:2013}
Alejandro Manjavacas, Federico Marchesin, Sukosin Thongrattanasiri, Peter
  Koval, Peter Nordlander, Daniel Sánchez-Portal, and F.~Javier García~de
  Abajo.
\newblock Tunable molecular plasmons in polycyclic aromatic hydrocarbons.
\newblock {\em ACS Nano}, 7(4):3635--3643, 2013.
\newblock PMID: 23484678.

\bibitem{deAbajoManjavacas:2013}
Alejandro Manjavacas, Sukosin Thongrattanasiri, and F.~Javier~García de~Abajo.
\newblock Plasmons driven by single electrons in graphene nanoislands.
\newblock {\em Nanophotonics}, 2(2):139 -- 151, 01 Apr. 2013.

\bibitem{Pybinding:2020}
Dean Moldovan, Misa Andelkovic, and Francois Peeters.
\newblock {pybinding v0.9.5: a Python package for tight- binding calculations},
  August 2020.
\newblock {This work was supported by the Flemish Science Foundation (FWO-Vl)
  and the Methusalem Funding of the Flemish Government.}

\bibitem{Nagao:2010}
Tadaaki Nagao, Gui Han, ChungVu Hoang, Jung-Sub Wi, Annemarie Pucci, Daniel
  Weber, Frank Neubrech, Vyacheslav~M Silkin, Dominik Enders, Osamu Saito, and
  Masud Rana.
\newblock Plasmons in nanoscale and atomic-scale systems.
\newblock {\em Science and Technology of Advanced Materials}, 11(5):054506,
  2010.

\bibitem{Naik:2013}
Gururaj~V. Naik, Vladimir~M. Shalaev, and Alexandra Boltasseva.
\newblock Alternative plasmonic materials: Beyond gold and silver.
\newblock {\em Advanced Materials}, 25(24):3264--3294, 2013.

\bibitem{Onida:2002}
Giovanni Onida, Lucia Reining, and Angel Rubio.
\newblock Electronic excitations: density-functional versus many-body
  green's-function approaches.
\newblock {\em Rev. Mod. Phys.}, 74:601--659, Jun 2002.

\bibitem{Runge:1984}
Erich Runge and E.~K.~U. Gross.
\newblock Density-functional theory for time-dependent systems.
\newblock {\em Phys. Rev. Lett.}, 52:997--1000, Mar 1984.

\bibitem{Sharma:2015}
Medha Sharma and M.A.H. Ahsan.
\newblock Organization of the hilbert space for exact diagonalization of
  hubbard model.
\newblock {\em Computer Physics Communications}, 193:19 -- 29, 2015.

\bibitem{Stefanucci:Book}
Gianluca Stefanucci and Robert van Leeuwen.
\newblock {\em Nonequilibrium Many-Body Theory of Quantum Systems: A Modern
  Introduction}.
\newblock Cambridge University Press, 2013.

\bibitem{Thongrattanasiri:2012}
Sukosin Thongrattanasiri, Alejandro Manjavacas, and F.~Javier García~de Abajo.
\newblock Quantum finite-size effects in graphene plasmons.
\newblock {\em ACS Nano}, 6(2):1766--1775, 2012.
\newblock PMID: 22217250.

\bibitem{Yazyev:2010}
Oleg~V Yazyev.
\newblock Emergence of magnetism in graphene materials and nanostructures.
\newblock {\em Reports on Progress in Physics}, 73(5):056501, apr 2010.

\end{thebibliography}


\begin{thebibliography}{1}

\bibitem{Pybinding:2020}
Dean Moldovan, Misa Andelkovic, and Francois Peeters.
\newblock {pybinding v0.9.5: a Python package for tight- binding calculations},
  August 2020.
\newblock {This work was supported by the Flemish Science Foundation (FWO-Vl)
  and the Methusalem Funding of the Flemish Government.}

\end{thebibliography}
\nocite{*}

\end{document}


\preprint{APS/123-QED}

\title{Supporting Information: Many-body formalism of optical absorption in nanosystems and molecules.}

\author{Antoine Honet}
\affiliation{%
Department of Physics and Namur Institute of Structured Materials, University of Namur, Rue de Bruxelles 51, 5000 Namur, Belgium
}%

\author{Luc Henrard}
\affiliation{%
Department of Physics and Namur Institute of Structured Materials, University of Namur, Rue de Bruxelles 51, 5000 Namur, Belgium
}%

\author{Vincent Meunier}%
\affiliation{%
 Department of Physics, Applied Physics, and
Astronomy, Rensselaer Polytechnic Institute, Troy, New York 12180, USA,
}%

\maketitle


\author{Antoine Honet, Luc Henrard and Vincent Meunier}

\date{\today}

\section{Numerical details.} The structures have been generated using the Pybinding software~\cite{Pybinding:2020}. A value of $\eta = 2.10^{-4}\frac{E_H}{t} \simeq 0.05 eV$ -- with $E_H=27.21 eV$ the Hartree energy -- have been used and $2^{15}$ values of frequency ranging from $-4\pi t$ to $4\pi t$ were employed to achieve numerical convergence and good peak description.

We use a mixing scheme to accelerate convergence of the Green's function: the starting retarded Green's function at iteration $k+1$ ($G^{R,in}_{k+1}$) is given by:
\begin{equation}
    G^{R,in}_{k+1} = \alpha G^{R, out}_k + (1-\alpha-\alpha')  G^{R, out}_{k-1} + \alpha' G^{R, out}_{k-2}
\end{equation}
where $G^{R, out}_k$ is the retarded Green's function calculated from the Dyson equation (see main text) at iteration $k$, and $\alpha$ and $\alpha'$ are mixing parameters.

In all numerical results presented here, the values $\alpha =0.8$ and $\alpha'=0$ were used as they were found to help the convergence scheme best. In the case of triphenylene, we used the values $\alpha =0.5$ and $\alpha'=0.25$, as those parameters were found to yield faster convergence.

The Fermi levels have been calculated and aligned after each iteration in order to conserve the total number of particles. The value $k_BT=25~meV$ was used throughout this paper.

In practice, Fast Fourier Transforms (FFT) have been used (numpy library) to convert information back and forth between time and frequency domains. 

\section{Role of the $\eta$ and $\zeta$ parameters and equivalence of the Lindhard formula and the Green's functions formalism.}

The Green's functions formulation of the non-interacting susceptibility (see Eq. (9) of the main text) is mathematically equivalent to the Lindhard formula (Eq. (3) of main text). This can be demonstrated by taking the Fourier transform of the time-domain non-interacting susceptibility in the Green's function formalism and then using the definition of the Fourier transform of the Green's functions and the integral expressions of the Heaviside function and Dirac delta function:
\begin{equation}
    \begin{split}
        &\Theta(t) = \frac{i}{2\pi}\lim_{\zeta\rightarrow 0} \int_{-\infty}^{+\infty} \frac{e^{-i\omega t}}{\omega+i\zeta} \dd \omega\\
        &2\pi \delta (\omega-\omega') = \int_{-\infty}^{+\infty} e^{i(\omega-\omega') t} \dd t
    \end{split}
\end{equation}
where $\zeta$ is a small positive parameter.

We also need to remember the expression of the non-interacting lesser and greater Green's functions to find back the Lindhard formula:
\begin{equation}
\begin{split}
     & G_{0,il\sigma,jl'\sigma}^<(\omega) = 2\pi i \sum_k f_{FD} (E_k) \delta(\omega-E_k) a_{il\sigma}^k (a_{jl'\sigma}^k)^*\\
     & G_{0,il\sigma,jl'\sigma}^>(\omega) = -2\pi i \sum_k \bar{f}_{FD}(E_k) \delta(\omega-E_k) a_{il\sigma}^k (a_{jl'\sigma}^k)^*.
\end{split}
\label{less_great}
\end{equation}

Numerically, the Dirac delta functions in the Green's functions are approximated by
\begin{equation}
    -\pi\delta(x) = \lim_{\eta\rightarrow 0} \Im(\frac{1}{x+i\eta}).
    \label{delta_eta}
\end{equation}
with a broadening parameter $\eta$.
Inserting Eq.(\ref{delta_eta}) into Eq.(\ref{less_great}) leads to:

\begin{equation}
    \begin{split}
    & G_{0,il\sigma,jl'\sigma}^<(\omega) =  \sum_k f_{FD} (E_k) \frac{2i\eta}{(\omega-E_k)^2+\eta^2} a_{il\sigma}^k (a_{jl'\sigma}^k)^* \\
     & G_{0,il\sigma,jl'\sigma}^>(\omega) = -\sum_k \bar{f}_{FD}(E_k) \frac{2i\eta}{(\omega-E_k)^2+\eta^2}a_{il\sigma}^k (a_{jl'\sigma}^k)^*
    \end{split}
    \label{less_great_eta}
\end{equation}
Eq.(\ref{less_great_eta}) is equivalent to the definitions in the main text in term of the matrix elements of the inverse of the Hamiltonian (Eqs. (7) and (8) in main text). 

\begin{figure}
\centering
    \includegraphics[width=.5\textwidth]{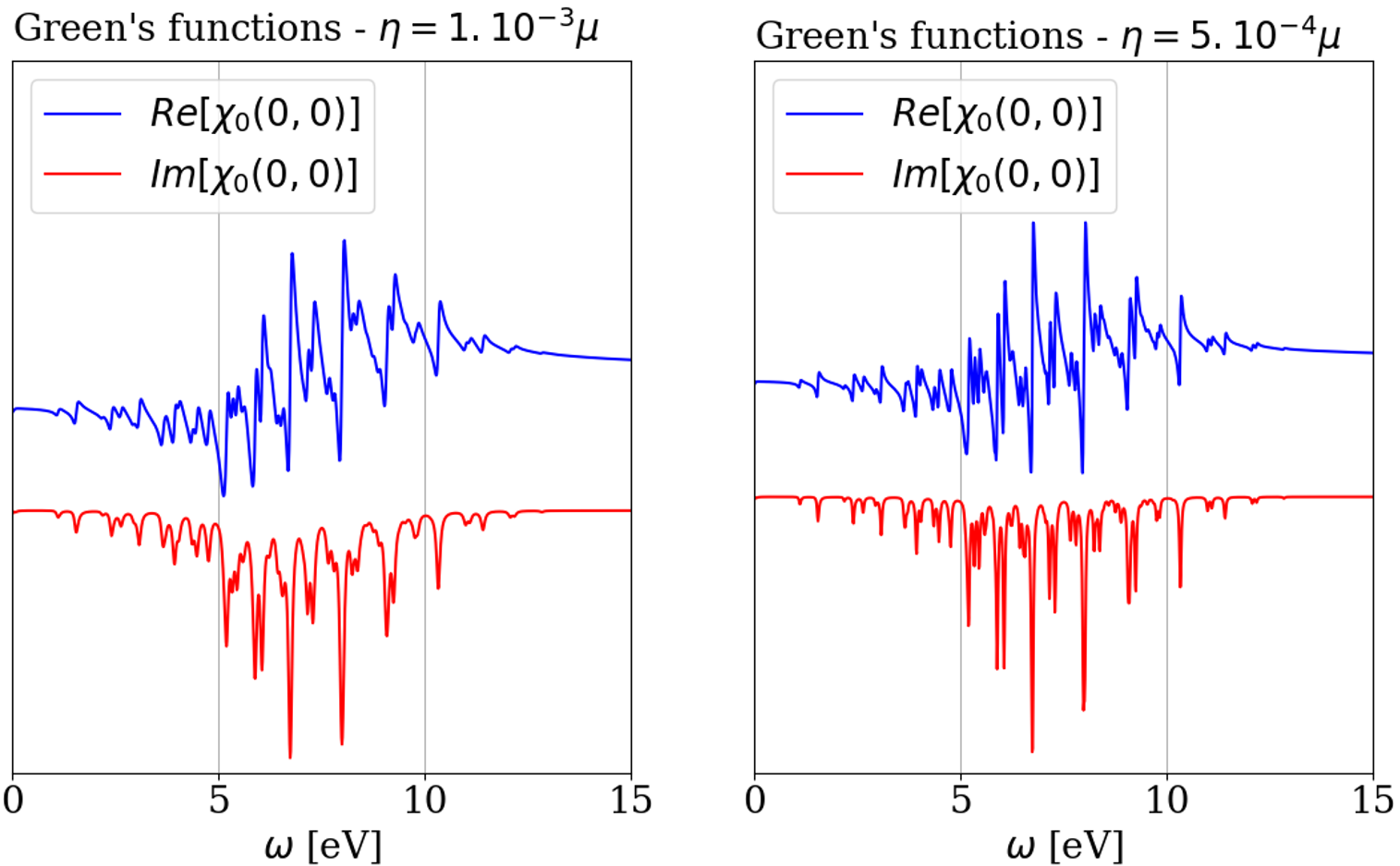}
\caption{Real and imaginary parts of the $(0,0)$-component of the non-interacting susceptibility $\chi_0(\vec{r},\vec{r}^\prime)$ for the tetracene molecule computed with the Green's function formalism applied to the TB model for $\eta = 10^{-3} \mu$ (left) and for $\eta = 5.10^{-4} \mu$ (right). $\mu = \frac{E_H}{t}$ with $E_H$ the Hartree energy and $t=2.6$~eV the hopping parameter. $\vec{r}=0$ corresponds to the atomic site located upwards at the extreme left of tetracene (See Fig. 1 of the main text).}
\label{fig:GF_eta}
\end{figure}

The $\zeta$ and $\eta$ parameters in the Lindhard formula and in the Green's functions formalism have a mathematically distinct origin (either from the definition of the Heaviside or the Dirac delta function) but they represent the same phenomenon: the natural broadening of the electronic transitions in the non-interacting susceptibility $\chi_0$, which is itself related to the life-time of the electron in the state. Fig.~\ref{fig:GF_eta} shows the influence of the $\eta$ parameter on $\chi_0$ computed in the Green's functions formalism for the TB model ($\chi_0(0,0)$ is given as illustration). Increasing the value of $\eta$ leads to a broadening of the peaks in the non-interacting $\chi_0$.

\begin{figure}
\centering
    \includegraphics[width=.5\textwidth]{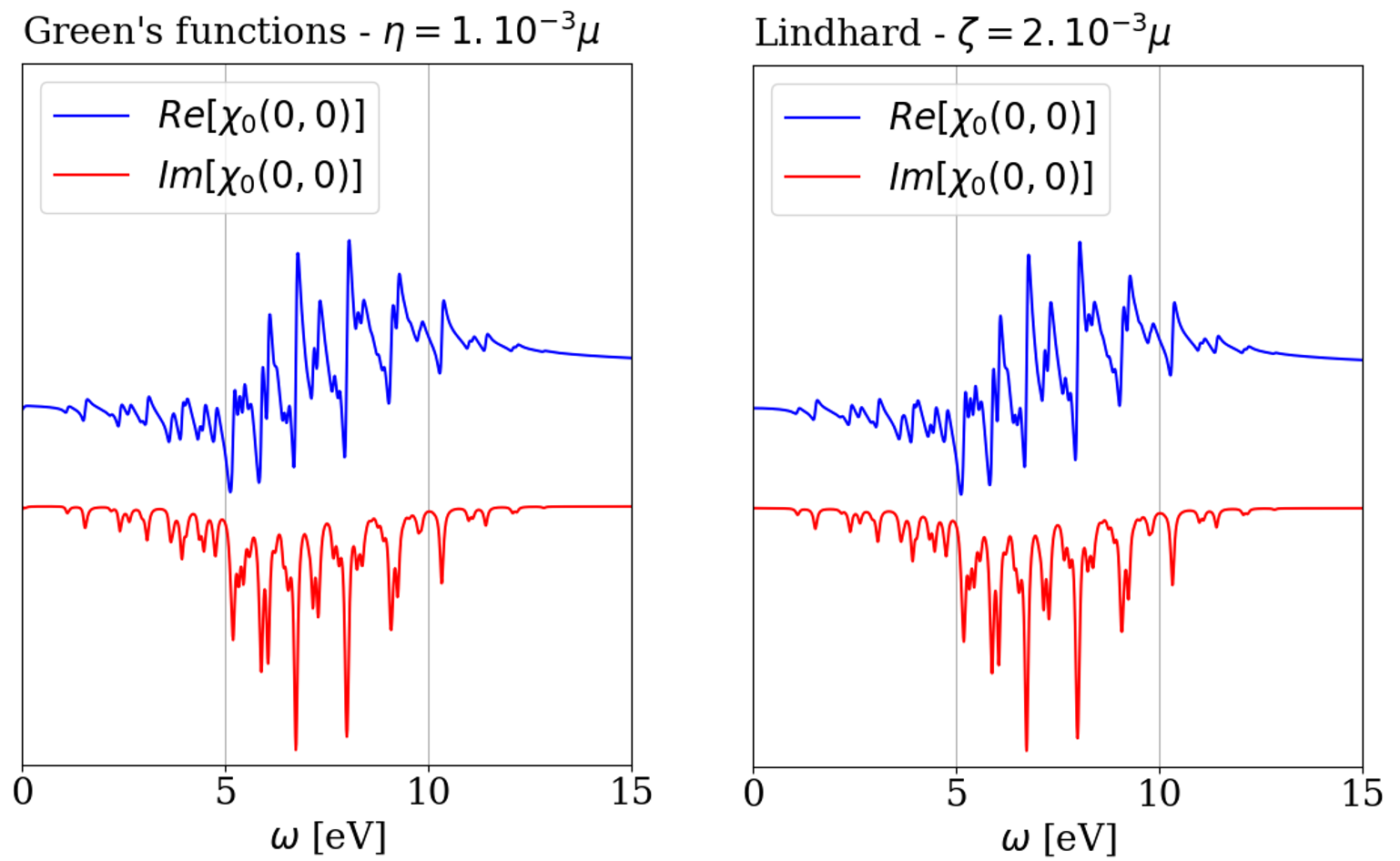}
\caption{Real and imaginary parts of the $(0,0)$-component of the non-interacting susceptibility $\chi_0(\vec{r},\vec{r}^\prime)$ for the TB approximation of the tetracene molecule computed with the Green's functions formalism for $\eta = 10^{-3} \mu$ (left) and with the Lindhard formula for $\zeta = 2.10^{-3} \mu$ (right). $\mu = \frac{E_H}{t}$ with $E_H$ the Hartree energy and $t=2.6$~eV the hopping parameter. $\vec{r}=0$ corresponds to the atomic site located upwards at the extreme left of tetracene (See Fig. 1 of the main text).}
\label{fig:GF_Lind}
\end{figure}

As stated before, $\zeta$ and $\eta$ stands for the same broadening mechanism but they differ from their mathematical origin and there is no obvious mathematical relation between them. Nevertheless, we found that a value of $\zeta$ twice as big as $\eta$ represents approximately the same broadening as can be seen from Fig.~\ref{fig:GF_Lind}. This figure also shows numerically the equivalence between the Green's functions approach and the Lindhard formula.

\section{Coulomb screening in RPA and polarization effects for the TB, MF-H and GW-H models.}

\begin{figure*}
\centering
    \includegraphics[width=\textwidth]{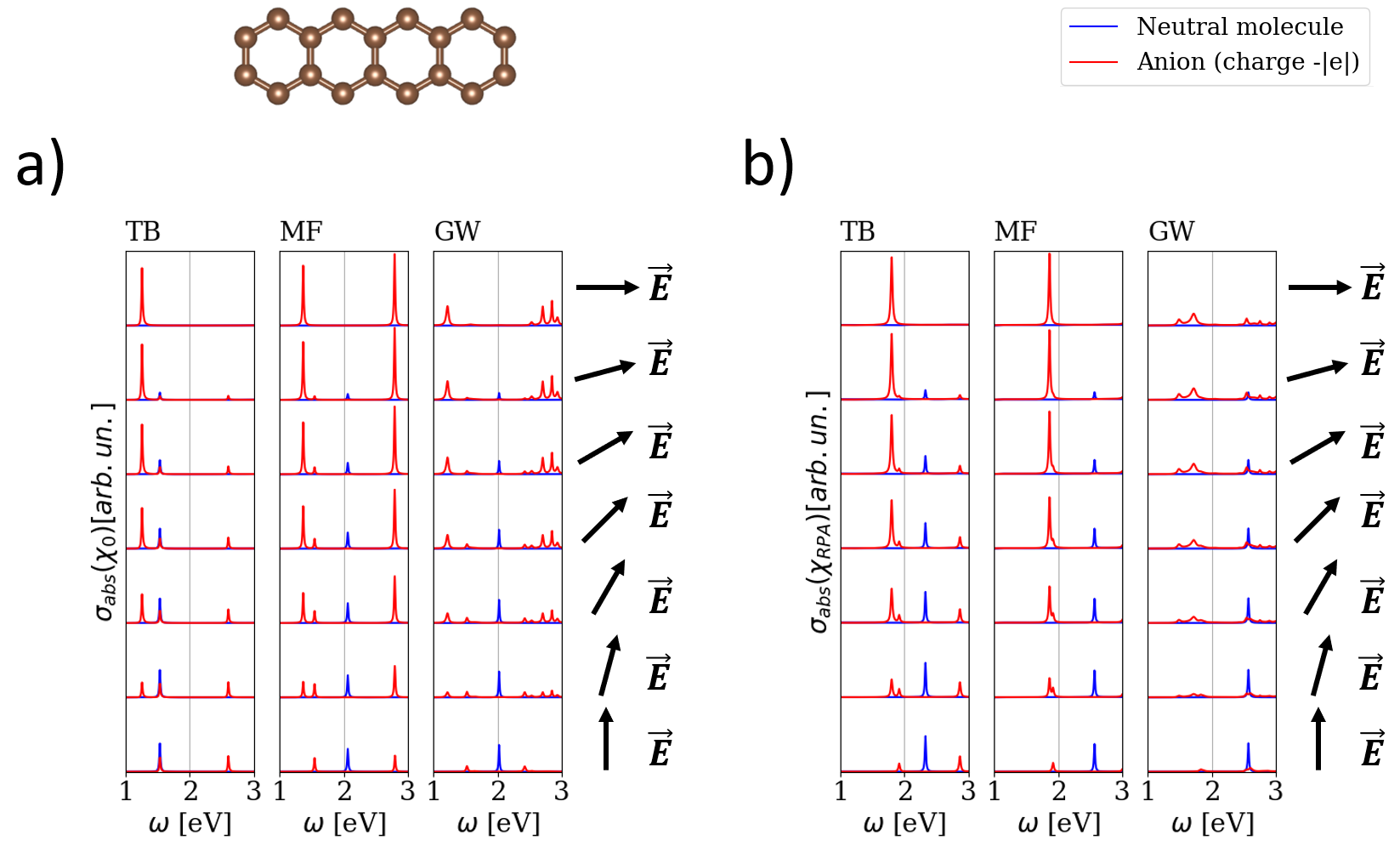}
\caption{Optical absorption cross section of the tetracene molecule in the neutral
state (blue curves) and in the one electron charged state (red curves). Absorption
cross sections are computed without taking into account the Coulomb screening
in RPA (from $\chi_{0}$) in (a) and with the screening (from $\chi_{RPA}$) in (b). In both
(a) and (b), the left figures are from the TB model, the middle figures are from
the MF-H model and the right figures are from the GW-H model.
In each graph, the curves correspond to different polarizations of the incident
electric field from electric field aligned with the main axis of the molecule(top curves) to
electric field perpendicular to them (bottom curves) (see arrows at the right of the graphs).
Angles are spaced by $\frac{\pi}{12}$.}
\label{fig:Full_compar}
\end{figure*}

Fig.~\ref{fig:Full_compar} shows the optical absorption cross section of the neutral and the single-electron charged tetracene in the three models considered (TB, MF-H and GW-H) for different orientations of the incident electric field and including or not the RPA screening.

The description of the Coulomb screening using the RPA is crucial (comparing fig.~\ref{fig:Full_compar} (a) and fig.~\ref{fig:Full_compar} (b)) in all the three models: in GW, the spin correlation is taken into account but the Coulomb screening still need to be included through RPA.

The atomic position of the molecules considered in this work lie in the same plane. It is clear from Eq. (6) of main text that there are no response for an electric field perpendicular to this plane.

\section{Role of the $U$ parameter in the MF-H model}
\begin{figure}
\centering
    \includegraphics[width=0.5\textwidth]{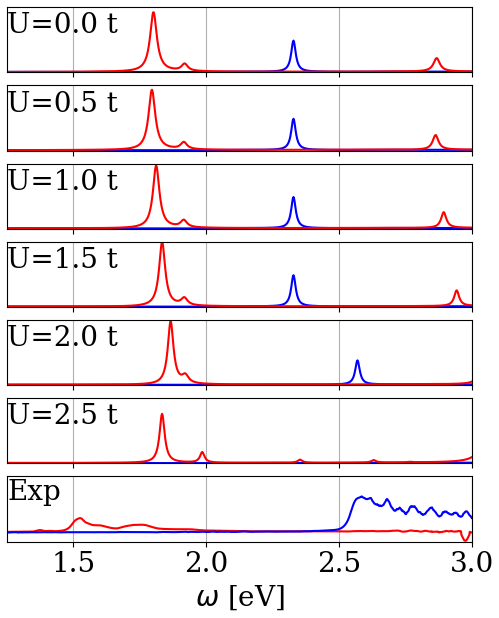}
\caption{Comparison of the optical absorption cross section between the and for the MF-H model for different values of $U$ ranging from $U=0$ to $U=2.5 t$ and the experimental results (at the bottom of the figure) for the tetracene molecule in the neutral state (blue curves) and in the one electron charged state (red curves). For the theoretical calculations, an average over 7 different angles is taken (see main text). The cross section are given in arbitrary units that are the same for the theoretical curves. The values of the parameters are $t=2.6 eV$ and $\eta = 2.10^{-4}\frac{E_H}{t}$.}
\label{fig:MF_U}
\end{figure}

In the main text, the role of the $U$ parameter has been investigated in the GW-H model, leading to a choice of the value $U=2t$. The results of the optical absorption cross section in the MF-H model for the tetracene molecule are shown at fig.\ref{fig:MF_U} for different values of $U$. The low-energy peaks of the charged particle present almost no shift while the $U$ parameter increases. The peak of the neutral particle undergoes a blueshift between $U=1.5t$ and $U=2t$ and occurs at an energy larger than $3$ eV for $U=2.5t$. None of the $U$ value reproduce satisfactorily the experimental data as well as in the GW-H model.

\bibliographystyle{plain}
\bibliography{references}